\documentclass[12pt]{article}
\usepackage[dvips]{graphicx}
\usepackage{amssymb}
\usepackage{amsmath,amsthm}

\usepackage{bm}
\bmdefine{\Bzero}{0}
\bmdefine{\Bone}{1}

\def\Bx{{\bf x}}
\def\By{{\bf y}}

\def\BY{{\bf Y}}

\def\Bc{{\bf c}}
\def\Bbeta{{\bf \beta}}

\newtheorem{theorem}{Theorem}[section]
\newtheorem{lemma}{Lemma}[section]
\newtheorem{corollary}{Corollary}[section]

\newtheorem{definition}{Definition}[section]
\newtheorem{proposition}{Proposition}[section]

\newtheorem{conjecture}{Conjecture}[section]
\title{Some characterizations of affinely full-dimensional factorial
designs}
\author{
Satoshi Aoki\footnote{Department of Mathematics and Computer Science, 
Kagoshima University}
\footnote{CREST, JST}
and
Akimichi Takemura\footnote{
Graduate School of Information Science and Technology, University of Tokyo}
\footnote{CREST, JST}
}

\date{December, 2008}
\begin{document}
\maketitle
\begin{abstract}
A new class of two-level non-regular fractional factorial designs is
 defined. 
We call this class an {\it affinely full-dimensional factorial design}, 
meaning that design points in the design of this class are not contained 
in any
affine hyperplane in the vector space over $\mathbb{F}_2$. 
The property of the indicator
 function for this class is also clarified. 
A fractional
factorial design in this class has a desirable property that parameters
of the main effect model are simultaneously identifiable. 
We investigate the property of this class from the viewpoint of
 $D$-optimality. In particular, for the saturated designs, the
 $D$-optimal design is chosen from this class for the run sizes $r \equiv
 5,6,7$ (mod $8$).

\end{abstract}

\section{Introduction}
In the literature on two-level fractional factorial designs, 
regular fractional factorial designs have been mainly studied 
both in theory and applications. The reason is that properly chosen 
regular fractional factorial designs have many desirable properties of
being balanced and orthogonal. In addition, the regular fractional
factorial designs are easily constructed and used based on 
important concepts such as resolution and aberration. 
An elegant theory based on the linear algebra over $\mathbb{F}_2$ is 
well established for regular two-level fractional factorial designs. See
\cite{Mukerjee-Wu-2006} for example. 

On the other hand, non-regular designs have also been receiving attention of
researchers over the years, in particular for some specific topics 
such as Plackett-Burman designs, Hall's designs and mixed-level
orthogonal arrays.
However, it is very difficult to derive theoretical results for {\it
general} non-regular fractional factorial designs. Suppose
we have $s$ controllable factors of two-levels and want to construct a
fractional
factorial design with $r$ runs, where $r$ is neither a $1/2$-, $1/4$-,
$1/8$-, $\ldots$ fraction 
of the $2^s$ full factorial design. 
One approach to choose an $r$-runs design is 
to rely on various optimal criteria such as $D$-, $A$- or 
$E$-optimality. However, 
the problem of characterizing these optimal designs for practically 
many values of $(s, r)$ is combinatorially very difficult. 
See  \cite{Moyssiadis-Kounias-1982} and \cite{Chadjipantelis-et-al-1987} for
examples of 
$D$-optimal saturated designs. These works, in addition to enormous literature
on Hadamard matrices such as \cite{Hedayat-Wallis-1978}, indicate the
difficulty of the characterization of the $D$-optimal designs and the development of simple
algorithms for obtaining them. For the other criteria such as 
$A$- or $E$-optimality, see \cite{Pukelsheim-2006}. Another
approach for the
problem of optimal selection is various extension of the minimum
aberration criterion
to non-regular designs. See \cite{Deng-Tang-1999} and \cite{Tang-Deng-1999} for
minimum $G_2$-aberration, \cite{Xu-Wu-2001} for generalized minimum aberration,
\cite{Xu-2003} for minimum moment aberration.


In this paper, we give a new approach for investigating general non-regular
fractional factorial designs from a theoretical viewpoint.
We define a class of non-regular designs, which is derived
naturally
from the argument of identifiability of parameters. 
There are some other works considering such a {\it classification} of designs.
\cite{Balakrishnan-Yang-2006} gives a
classification of fractional factorial
designs with simple structure in view of their {\it indicator
functions}. The indicator function, first introduced
in \cite{Fontana-et-al-2000}, has become a powerful tool for studying general
non-regular
fractional factorial designs. See \cite{Cheng-et-al-2004} and
\cite{Loeppky-et-al-2006} 
for example. Since our work is motivated by the indicator
function approach, 
we also consider properties of the indicator function of the designs in 
our proposed class.

The construction of this paper is as follows. 
In Section 2, we define a new class of the two-level non-regular
fractional factorial designs and investigate its property. 
In Section 3, we consider relations between the class and $D$-optimal
designs. In particular we pay special attention to the saturated designs.
Finally in Section 4, we give some discussion.

\section{Definition of an affinely full-dimensional factorial design}
Suppose there are $s$ controllable factors of two levels. Let ${\cal D}$ be the
$2^s$ full factorial design with levels being $-1$ and $1$. ${\cal D}$
is written as
\[
 {\cal D}  = \{-1,1\}^s
 = \{(x_1,\ldots,x_s)\ |\ x_1^2 = \cdots = x_s^2 = 1\}.
\]
A fractional factorial design ${\cal F}$ (without replication) is a
subset of ${\cal D}$. Let
$r$ be the run size of ${\cal F}$. Therefore ${\cal F}$ is a set of 
$r$ points in ${\cal D}$. Arranging the elements of ${\cal F}$ appropriately, 
we represent ${\cal F}$ as an $r\times (s+1)$ matrix $M \in
\{-1,1\}^{r\times(s+1)}$
\begin{equation}
 M = \left(\begin{array}{cccc}
1 & m_{11} & \cdots & m_{1s}\\
\vdots & \vdots & & \vdots\\
1 & m_{r1} & \cdots & m_{rs}\\
\end{array}
\right), 
\label{eqn:def-of-M}
\end{equation}
where $m_{ij}$ is the level of the $j$th factor in the $i$th run. 
We call the left-most column of $M$ the ``$0$-th column'' and
write its elements as $m_{i0}\equiv 1, i = 1,\ldots,r$.
Note that $M$ is the design matrix of ${\cal F}$ for the main effect
model. If we write $\beta_j$ as the parameter of the main effect of the
$j$th factor for $j = 1,\ldots,s$, the linear model for the
observation variable $\BY = (Y_1,\ldots,Y_r)'$ is written as
\begin{equation}
\begin{array}{c}
 \BY = M\beta + \varepsilon,\ \varepsilon \sim N_m(\Bzero, \sigma^2I),\\
\Bbeta = (\beta_0,\beta_1,\ldots,\beta_s)',
\end{array}
\label{eqn:linear-model}
\end{equation} 
where $\beta_0$ is the parameter for intercept.
We put $m_{1j} = 1$ for $j = 1,\ldots,s$ without loss of generality
(by relabeling the two levels of each factor). 
We also assume that each column 
$(m_{1j},\ldots,m_{rj})'$, $j=1,\ldots,s$, 
contains at least one $-1$.

The indicator function $f$ of ${\cal F}$ is defined in
\cite{Fontana-et-al-2000} as a
function on ${\cal D}$ such that
\[
 f(\Bx) = \left\{\begin{array}{cl}
1, & \mbox{if}\ \Bx \in {\cal F},\\
0, & \mbox{if}\ \Bx \in {\cal D} \setminus {\cal F}.
\end{array}
\right.
\]
Following \cite{Fontana-et-al-2000}, 
we define contrasts $X_I(\Bx) = \prod_{i \in I}x_i$ on ${\cal D}$ for $I
\in {\cal P}$, where ${\cal P}$ is the set of all subsets of
$\{1,\ldots,s\}$. A fundamental fact in the theory of the experimental
design is that $\{X_I,\ I\in {\cal P}\}$ forms an orthogonal basis of
the set of all real-valued functions on ${\cal D}$. 
The indicator function $f$ of ${\cal F}$ is then written as the
polynomial form
\begin{equation}
 f(\Bx) = \sum_{I \in {\cal P}}b_IX_I(\Bx).
\label{eqn:indicator-func}
\end{equation}
Since $x_i^2 = 1$ on ${\cal D}$ for $i = 1,\ldots,s$, the above
expression is square-free and is unique on ${\cal D}$.

A regular $2^{s-k}$ fractional factorial design ${\cal A} \subset {\cal
D}$ is generated by $k$ linearly independent generating relations
\begin{equation}
\label{eq:defining-contrast}
 X_{I_1}(\Bx) = 1,\ldots, X_{I_k}(\Bx) = 1, 
\end{equation}
i.e.\ ${\cal A}=\{ \Bx \in {\cal D} \mid X_{I_{\ell}}(\Bx) = 1,\ \ell = 1,\ldots,k\}$.
${\cal A}$ contains $2^{s-k}$ points.
Note that the right hand sides of the relations
(\ref{eq:defining-contrast}) reflect the assumption $m_{1j} = 1$ for $j
= 1,\ldots,s$. In general, 
we can take $X_{I_{\ell}}(\Bx) = -1$ instead of
$X_{I_{\ell}}(\Bx) = 1$ in (\ref{eq:defining-contrast}). 
This just depends on the labeling of two levels
for each factor.  From randomization viewpoint, given the labelings of
two levels of each factor, it is desirable to choose $1$ or $-1$ randomly for
each generating relation in (\ref{eq:defining-contrast}).

Now we define a class of non-regular fractional factorial designs.
\begin{definition}
A non-regular fractional factorial design ${\cal F}$ is called an affinely
 full-dimensional factorial design if there is no regular fractional
 factorial design ${\cal A}$ satisfying ${\cal F}
 \subsetneq {\cal A}$. Conversely, 
a non-regular fractional factorial design ${\cal F}$ is called a subset
 fractional factorial design if there is some regular fractional
 factorial design ${\cal A}$ satisfying ${\cal F} \subsetneq {\cal A}$.
\end{definition}
The above definition gives a new class of the non-regular fractional
factorial designs. 
Any fractional factorial design with $r > 2^{s-1} $ is
an affinely full-dimensional factorial design.
The merit of the definition might not be clear at a glance. 
One of the properties of the affinely full-dimensional factorial design is the
simultaneous identifiability of the parameters in the model
(\ref{eqn:linear-model}).
Note that the least squares estimator of $\Bbeta$ is written as
$\hat{\Bbeta} = (M'M)^{-1}M'\By$ for the observation $\By$. We call
$\Bbeta$ is simultaneously identifiable if $M'M$ is a non-singular
matrix. We also note that, for regular fractional factorial designs, the
singularity of $M'M$ simply corresponds to a confounding relation of the main
effects. On the other hand, for the non-regular designs, 
the relation is not obvious. 
We clarify this point in the following lemma.
\begin{lemma}
\label{lem:1}
If ${\cal F}$ is an affinely full-dimensional factorial design, then
 $M'M$ is non-singular, i.e., $\Bbeta$ is simultaneously identifiable.
\end{lemma}
\paragraph*{Proof.}\ We show the contraposition. Suppose the columns of
 $M$ are linearly dependent. This dependence relation is preserved by the
 following operation: subtract the left-most 
column $(1,\dots, 1)'$ of $M$ from the other
 $s$ columns and then divide the $s$ columns 
 by $-2$.  By this operation, $1$ is mapped to $0$ and $-1$ is mapped to 
$1$.  
 Denote the resulting matrix  by $\widetilde 
 M=\{\widetilde m_{ij} \}$. Then $\widetilde m_{ij}=(1-m_{ij})/2$ for $j\ge 1$
 and $\widetilde m_{i0}\equiv 1$ for $i = 1,\ldots,r$.
 From the assumption of linearly dependence there 
 exists some integer vector $\Bc = (c_0,c_1,\ldots,c_s)'$ satisfying 
\begin{equation}
 \widetilde M\Bc = \Bzero. 
\label{eqn:proof-lemma1-depend}
\end{equation}
Considering the modulo $2$ reduction of (\ref{eqn:proof-lemma1-depend}), we have
\begin{equation} 
\widetilde M\Bc = \Bzero\ (\mbox{mod}\ 2),
\label{eqn:proof-lemma1-mod2}
\end{equation}
where the odd elements and the even elements of $\Bc$ are replaced by
$1$ and $0$, respectively. Here we can assume that there exists an odd
element of $\Bc$
in (\ref{eqn:proof-lemma1-depend}), since if every element of $\Bc$
is even then we can divide (\ref{eqn:proof-lemma1-depend}) 
by the power of $2$ in the factorization of the 
greatest common divisor of the elements of $\Bc$.
Moreover $c_0=0$, since the first row of $\widetilde M$ is
$(1,0,\dots,0)$. Then 
(\ref{eqn:proof-lemma1-mod2}) implies that there are even $1$'s
in $\{\widetilde m_{ij}\ |\ c_j = 1, j\ge 1\}$ for $i = 1,\ldots,r$,
or equivalently there are even $-1$'s in
$\{m_{ij}\ |\ c_j = 1, j\ge 1\}$ for $i = 1,\ldots,r$.  Therefore
$X_{I}(\Bx) = 1$ holds for 
$I = \{j\ |\ c_j = 1, j\ge 1\}$.
\hspace*{\fill}Q.E.D.

\ \\

\noindent
From this lemma, we also have the following corollary.
\begin{corollary}
Let $M$ be a design matrix of a  non-regular fractional factorial
 design. If $M'M$ is singular, then the design is a subset
 fractional factorial  design.
\end{corollary}

The proof of Lemma \ref{lem:1} clarifies the geometrical meaning of
the class.  By 
the correspondence $(1,-1)\leftrightarrow (0,1)$, 
$X_I(x_1,\ldots,x_s) = 1$, $(x_1,\dots, x_s)\in {\cal D}$ 
if and only if 
\[
 c_1 \widetilde x_1 + \cdots +  c_s \widetilde x_s = 0\ (\mbox{mod}\
 2),\quad
\widetilde x_j=(1-x_j)/2, \ j=1,\dots,s,
\]
where
\[
c_j = \left\{
 \begin{array}{cl}
1, & \mbox{if}\ j \in I,\\
0, & \mbox{if}\ j \not\in I.
\end{array}
\right.
\]
However, as remarked earlier, design points can be chosen by 
$X_I(x_1,\ldots,x_s) = -1$. 
Therefore, in general  there may  be a constant
term $c_0=0\  \text{or} \ 1$:
\begin{equation}
c_0+  c_1 \widetilde x_1 + \cdots +  c_s \widetilde x_s = 0\ (\mbox{mod}\
 2),\qquad
\widetilde x_j=(1-x_j)/2, \ j=1,\dots,s.
\label{eqn:affine-hyperplane}
\end{equation}
In this sense, 
${\cal F}$
is a  proper subset of no regular fractional factorial design if and only if
the points of $\cal F$,  considered as a subset of $\mathbb{F}_2^s =
\{0,1\}^s$, are not contained in any affine hyperplane of the form
(\ref{eqn:affine-hyperplane}), i.e., we can form a basis of
$\mathbb{F}_2^s = \{0,1\}^s$ as the
differences of the vectors in ${\cal F}$. 
This is the reason we call
the class {\it affinely full-dimensional}.

The question of determining whether a  given design is 
an affinely full-dimensional 
is  immediately read off
from its indicator function.
\begin{lemma}
Let ${\cal F}$ be a fractional factorial design and 
(\ref{eqn:indicator-func}) be its indicator function. Then ${\cal F}$ is
 an affinely full-dimensional factorial design if and only if
$
 |b_{I}| < b_{\emptyset}\ \mbox{for all}\ I\in {\cal P}.
$
\end{lemma}
\paragraph*{Proof.}\ Obvious from Proposition 4.2 and Corollary 4.3 of
\cite{Fontana-et-al-2000}.\hspace*{\fill}Q.E.D.

\section{$D$-optimality of the affinely full-dimensional factorial designs}
In Section 2, we define a new class of the non-regular fractional
factorial designs, namely, affinely full-dimensional factorial
designs. The affinely full-dimensional factorial designs have a
desirable property
that all the parameters are always simultaneously identifiable in the
saturated model. Next problem of interest is whether this class includes
good designs or not in view of various optimality criteria. In this
paper, we consider $D$-optimality of the designs. For given $r$ and $s$,
the $D$-optimal design is the matrix $M \in \{-1,1\}^{r\times (s+1)}$
which maximizes $|\det(M'M)|$. The $D$-optimal designs minimize the
generalized variance of $\hat{\beta}$ (\cite{Pukelsheim-2006}).

\subsection{$D$-optimal designs for the saturated cases}
First we consider the saturated cases, i.e., the cases of $r = s+1$. In
this case, the maximization of $|\det(M'M)|$ reduces to the maximization
of $|\det M|,\ M\in \{-1,1\}^{r\times r}$. This problem is known as the
{\it Hadamard maximal determinant problem}. 
Despite a century of works by mathematicians, this problem remains
unanswered in general.

To investigate the relation between the maximal determinant problem and
the affinely full-dimensional factorial designs, we show a basic 
theorem.
\begin{theorem}
A design  with a design matrix $M$ is affinely full-dimensional
 factorial if and only if $\det M$ is not divisible by $2^r$.
\label{thm:M-multiple-2^r}
\end{theorem}

We define $M$ and $\widetilde M$ as in the proof of Lemma \ref{lem:1}.
The columns of these matrices are numbered from $0$ to $s=r-1$ and
the rows are numbered from $1$ to $r$.
Note that the first row of $\widetilde{M}$ is $(1,0,\ldots,0)$.
This matrix has the following property.
\begin{lemma}
$\det\widetilde{M}$ is an odd integer if and only if $M$ is a design
 matrix of an affinely full-dimensional factorial design. 
\label{lemma:det-M-tilde}
\end{lemma}
\paragraph*{Proof of Lemma \ref{lemma:det-M-tilde}.}\ If $M$ is a design
 matrix of a subset fractional factorial design or a regular fractional
 factorial design, there exists $1 \leq j_1 < \cdots < j_p \leq s,\ p
 \geq 2$ satisfying
\[
 \prod_{\ell = 1}^p m_{ij_{\ell}} = 1
\]
for $i = 1,\ldots,r$. From the definition of $\widetilde{M}$, this is 
equivalent to 
\[
 \sum_{\ell = 1}^p \widetilde{m}_{ij_{\ell}} = 0\ \ {({\rm mod}\ 2)}
\]
for $i = 1,\ldots,r$. It means that $\widetilde{M}$ is a singular matrix
in ${\mathbb F}_2$, i.e., $\det\widetilde{M} = 0\ \ ({\rm mod}\
2)$.\\
\hspace*{\fill}Q.E.D.

\paragraph*{Proof of Theorem \ref{thm:M-multiple-2^r}.}\ From the
definition, we construct $\widetilde{M}$ from $M$ as follows:
\begin{itemize}
\item subtracting the left-most column of $M$ from the other columns,
\[
 m_{ij} \leftarrow m_{ij} - m_{i0},\ i = 1,\ldots,r,\ j = 1,\ldots,s,
\]
\item dividing the columns except for the left-most column by $-2$,
\[
 m_{ij} \leftarrow m_{ij} / (-2), \ i = 1,\ldots,r,\ j = 1,\ldots,s.
\] 
\end{itemize}
Note that, after the first operation, no column of $M$ is $(0,\ldots,0)'$
from our assumption. Therefore $(-2)^{r-1}\det \widetilde{M} = \det M$ holds.
From Lemma \ref{lemma:det-M-tilde}, we have proved the theorem.\\
\hspace*{\fill}Q.E.D.

\ \\

\noindent
Theorem \ref{thm:M-multiple-2^r} shows that whether a given
non-regular fractional factorial design is affinely
full-dimensional or not is judged from its determinant. 
Using this characteristic, we investigate the $D$-optimal designs.

There are many literature reporting the solution of the Hadamard maximal
determinant problem for specific $r$. The most basic result
was given by Hadamard (\cite{Hadamard-1893}) as $\det M \leq r^{r/2}$,
where the bound is achieved only for Hadamard matrices. Paley has
conjectured that a Hadamard matrix exists for every $r = 0\ ({\rm mod}\
4)$. The lowest order for which a Hadamard matrix is not yet
obtained is $r = 668$ (\cite{Seberry-Yamada-1992}). For a design
where the design matrix $M$ is a Hadamard matrix, it is a subset
fractional factorial design since the bound reduces to $(4k)^{r/2} = 2^r k^{r/2}$.
In fact, the product of the elements in each
row is $1$ from the property of Hadamard matrix. 
Therefore we consider the case that $r$ is not divisible by
$4$. According to the literature reporting the maximal determinant
matrices, we investigate whether the $D$-optimal design matrix is 
affinely full-dimensional factorial or not for $r < 100$. 
The results are given in
Table \ref{tbl:table-of-maximal-determinant}.
\begin{table}[htbp]
\begin{center}
\caption{Summary whether the maximal determinant matrices are affinely
 full-dimensional factorial design (Yes) or subset fractional factorial
 design (No) for $r = 4,\ldots,99$. The values of the maximal determinant are given as
divided by $2^{r-1}$. The values that achieve the bounds (\cite{Barba-1933}, \cite{Ehlich-1964}, 
\cite{Wojtas-1964} and \cite{Ehlich-1964b}) are underlined.}\  \\
\label{tbl:table-of-maximal-determinant}
\begin{tabular}{cc|ccc|ccc|ccc}
$r$ & class & $r$ & class & Det$/2^{r-1}$ & $r$ & class & Det$/2^{r-1}$
 & $r$ & class
 & Det$/2^{r-1}$ \\ \hline
$4$  & No &$5$ & Yes & \underline{$3$} & $6$ & Yes & \underline{$5$} & $7$ & Yes & \underline{$9$}\\
$8$  & No &$9$  & No  & $7\cdot 2^3$&$10$ & No  & \underline{$9\cdot 2^4$}&$11$ &
 No & $2^{6}\cdot 5$\\
$12$ & No &$13$ & Yes & \underline{$5\cdot 3^6$}&$14$ & Yes & \underline{$13\cdot 3^6$} &$15$ &
 Yes& $3^6\cdot 5\cdot 7$\\
$16$ & No &$17$ & No  & $2^{16}\cdot 5$&$18$ & No  & \underline{$17\cdot 2^8\cdot 2^8$} 
&$19$ & ?  &\\
$20$ & No &$21$ & Yes & $5^9\cdot 29$&$22$ & ?   & &$23$ & ?  &\\
$24$ & No &$25$ & No  & \underline{$7\cdot 6^{12}$}&$26$ & No &
 \underline{$25 \cdot 3^{12} \cdot 2^{12}$} &$27$ & ?  &\\
$28$ & No &$29$ & ?   & &$30$ & Yes & \underline{$29\cdot 7^{14}$} &$31$ & ?  &\\
$32$ & No &$33$ & ?   & &$34$ & ?   & &$35$ & ?  &\\
$36$ & No &$37$ & ?   & &$38$ & Yes & \underline{$37\cdot 9^{18}$} &$39$ & ?  &\\
$40$ & No &$41$ & No  & \underline{$9\cdot 10^{20}$} &$42$ & No &
 \underline{$41 \cdot 5^{20}\cdot 2^{20}$} &$43$
 & ?  &\\
$44$ & No &$45$ & ?   & &$46$ & Yes & \underline{$45\cdot 11^{22}$} &$47$ & ?  &\\
$48$ & No &$49$ & ?   & &$50$ & No  & \underline{$49\cdot 6^{24}\cdot 2^{24}$}
 &$51$ & ?  &\\
$52$ & No &$53$ & ?   & &$54$ & Yes & \underline{$53\cdot 13^{26}$} &$55$ & ?  &\\
$56$ & No &$57$ & ?   & &$58$ & ?   & &$59$ & ?  &\\
$60$ & No &$61$ & Yes & \underline{$11\cdot 15^{30}$}&$62$ & Yes &
 \underline{$61\cdot 15^{30}$} &$63$ & ?  &\\
$64$ & No &$65$ & ?   & &$66$ & No  & \underline{$65\cdot 8^{32} \cdot 2^{32}$} &$67$ & ?  &\\
$68$ & No &$69$ & ?   & &$70$ & ?   & &$71$ & ?  &\\
$72$ & No &$73$ & ?   & &$74$ & No  & \underline{$73\cdot 9^{36}\cdot 2^{36}$} &$75$
 & ?  &\\
$76$ & No &$77$ & ?   & &$78$ & ?   & &$79$ & ?  &\\
$80$ & No &$81$ & ?   & &$82$ & No  & \underline{$81\cdot {10}^{40}\cdot 2^{40}$} &$83$
 & ?  &\\
$84$ & No &$85$ & ?   & &$86$ & Yes & \underline{$85\cdot 21^{42}$} &$87$ & ?  &\\
$88$ & No &$89$ & ?   & &$90$ & No  & \underline{$89\cdot 11^{44}\cdot 2^{44}$} &$91$
 & ?  &\\
$92$ & No &$93$ & ?   & &$94$ & ?   & &$95$ & ?  &\\
$96$ & No &$97$ & ?   & &$98$ & No  & \underline{$97\cdot 12^{48}\cdot 2^{48}$} &$99$
 & ?  &\\
\end{tabular}
\end{center}
\end{table}
In Table \ref{tbl:table-of-maximal-determinant}, ``?'' means that the
maximal determinant matrix is not obtained (or the maximality is not proved) at present. We also show the values of the maximal
determinant divided by $2^{r-1}$ for $r = 1,2,3$ (mod $4$).
We underline the values, when 
the values achieve the known bounds such as \cite{Barba-1933} for 
$r\equiv 1$ (mod $4$), \cite{Ehlich-1964} and \cite{Wojtas-1964} for 
$r\equiv 2$ (mod $4$) and \cite{Ehlich-1964b} for $r\equiv 3$ (mod $4$). 
We omit the reference for the Hadamard
matrices. For the other maximal determinant matrices, see 
\cite{Mood-1946} for $r = 5$; 
\cite{Williamson-1946} for $r = 6, 7$; 
\cite{Ehlich-Zeller-1962} for $r = 9, 10,11$;
\cite{Smith-1988} for $r = 9, 10, 15$;
\cite{Wojtas-1964} for $r = 10, 26$; 
\cite{Galil-Kiefer-1980} for $r = 11$; 
\cite{Raghavarao-1959} for $r = 13, 25, 50$; 
\cite{Colbourn-Dinitz-1996} for $r = 13$; 
\cite{Ehlich-1964} for $r = 14, 18, 26, 30, 38$; 
\cite{Cohn-1994} for $r = 14, 18$; 
\cite{Cohn-1989} for $r = 15, 74, 82, 90, 98$; 
\cite{Cohn-2000} and \cite{Orrick-2005} for $r = 15$; 
\cite{Schmidt-1973}, \cite{Moyssiadis-Kounias-1982} and \cite{Husain-1945} for $r = 17$; 
\cite{Yang-1968} for $r = 18, 26, 30, 38$; 
\cite{Chadjipantelis-et-al-1987} and \cite{Orrick-2008} for $r = 21$; 
\cite{Brouwer-1983} for $r = 25, 61$; 
\cite{Kounias-et-al-1994} for $r = 26, 30, 38, 42, 46, 50, 54, 66$; 
\cite{Bridges-Hall-Hayden-1981}, \cite{van-Trung-1982}, \cite{Spence-1993} and 
\cite{Spence-1995} for $r = 41$; 
\cite{Yang-1966a} and \cite{Kharaghani-1987} for $r = 42, 66$; 
\cite{Yang-1969} for $r = 50, 62$; 
\cite{Yang-1966b} for $r = 54$; 
\cite{Pavcevic-Spence-1999a} and \cite{Pavcevic-Spence-1999b} for $r = 61$; 
\cite{Yang-1976} for $r = 66$; 
\cite{Chadjipantelis-Kounias-1985} for $r = 86$; 
\cite{Kounias-et-al-1996} for $r = 90$.

Table \ref{tbl:table-of-maximal-determinant} shows an interesting
periodicity. We present the following conjecture.
\begin{conjecture}
For $r = 5,6,7$ (mod $8$), $D$-optimal design is affinely full-dimensional
factorial. For $r = 0,1,2,3,4$ (mod $8$), $D$-optimal design is 
subset fractional factorial.
\label{conj:mod-8-rule}
\end{conjecture}

Though the authors do not succeed in proving this conjecture,
we do not find any counterexample at present.
We consider special cases instead.
For the case $r = 1$ (mod $4$),
\cite{Barba-1933} gives a bound
\begin{equation}
\det M \leq (2r - 1)^{1/2} (r - 1)^{(r - 1)/2}.
\label{eqn:Barba-bound}
\end{equation}
See also \cite{Ehlich-1964} and \cite{Wojtas-1964}.
This bound is achievable only if $2r - 1$ is a perfect square.
In fact, 
for $r = 5, 13, 25, 41, 61$ in Table \ref{tbl:table-of-maximal-determinant},
the maximum determinant achieves the bound, whereas 
the maximal determinant matrix has not been founded at present for $r =
85$. For the cases that the bound
is attained, it is easy to prove the conjecture.
\begin{proposition}
For $r = 1$ (mod $8$) and the bound (\ref{eqn:Barba-bound}) is attained,
the maximal determinant matrix is chosen as a design matrix for the
 subset fractional
factorial design. 
On the other hand, for $r = 5$ (mod $8$) and the bound
 (\ref{eqn:Barba-bound}) is attained,
the maximal determinant matrix is chosen as a design matrix for the affinely
 full-dimensional factorial design. 
\end{proposition}
\paragraph*{Proof.}\ Write $2r - 1 = m^2$ and $r = 4k + 1$. Then it holds
\[
(2r - 1)^{1/2} (r - 1)^{(r - 1)/2} = m 2^{r-1} k^{2k}.
\]
Since $m$ is an odd integer, the maximal determinant is divisible by $2^r$
if and only if $k$ is an even integer, i.e., $r = 1$ (mod
$8$). \hspace*{\fill}Q.E.D.

\ \\
\noindent
Similarly, for the case that $r=2$ (mod $4$), \cite{Ehlich-1964} and
\cite{Wojtas-1964} give a bound
\begin{equation}
 \det M \leq 2(r-1)(r-2)^{(r-2)/2}.
\label{eqn:Ehlich-Wojtas-bound}
\end{equation}
In Table \ref{tbl:table-of-maximal-determinant}, all the maximum
determinants achieve the bound for $r = 6,10,\ldots,98$ except for
unsolved cases, $r = 22, 34, 58, 70, 78, 94$. For the cases that the
bound is achieved, it is also easy to prove the conjecture.
\begin{proposition}
For $r = 2$ (mod $8$) and the bound (\ref{eqn:Ehlich-Wojtas-bound}) is
 attained,
the maximal determinant matrix is chosen as a design matrix for the
 subset fractional
factorial design. 
On the other hand, for $r = 6$ (mod $8$) and the bound
 (\ref{eqn:Ehlich-Wojtas-bound}) is attained,
the maximal determinant matrix is chosen as a design matrix for the affinely
 full-dimensional
factorial design. 
\end{proposition}
\paragraph*{Proof.}\ For $r = 8k + 2, k \geq 1$, the bound, 
\[
2(r-1)(r-2)^{(r-2)/2} = (r-1)k^{4k}2^{r+4k-1}, 
\]
is divisible by $2^r$ since $k \geq 1$. For $r = 8k + 6, k \geq 0$, the
bound, 
\[
2(r-1)(r-2)^{(r-2)/2} = (r-1)(2k + 1)^{4k + 2}2^{r-1}, 
\]
is not divisible by $2^r$ since both $r-1$ and $2k+1$ are odd
integers. \hspace*{\fill}Q.E.D.

\ \\
\noindent
For the cases that $r = 3$ (mod $4$), though a similar bound is given by 
\cite{Ehlich-1964b}, the maximal determinant matrix attaining the bound
is not found at present except for $r = 3$.

\subsection{$D$-optimal designs for $4$ and $5$ factors cases}
Next we consider the non-saturated cases of $s = 4,5$. We enumerate all
the design matrices of the runs $r \leq 10$ and obtain the $D$-optimal
designs. We investigate whether the $D$-optimal designs are the affinely
full-dimensional factorial design or the subset fractional factorial
design. The result is summarized in 
Table \ref{tbl:non-saturated-s=4} and \ref{tbl:non-saturated-s=5}.
Note that 
the factorial design is always affinely full-dimensional for $r > 8, s =
4$ since $r > 2^{s-1}$. 
\begin{table}[htbp]
\begin{center}
\caption{$D$-optimal designs for $4$ factors}
\label{tbl:non-saturated-s=4}
\begin{tabular}{lccc}\hline
$r$    & $5$ & $6$ & $7$\\
{\begin{tabular}{l}
$D$-optimal\\
design 
\end{tabular}}
& 
${
\begin{array}{rrrr}
x_1 & x_2 & x_3 & x_4\\ \hline
1 & 1 & 1 & 1\\
1 & 1 &-1 &-1\\
1 &-1 & 1 &-1\\
-1 & 1 & 1 &-1\\
-1 &-1 &-1 & 1\\
& & & \\
& & & 
\end{array}
}$ & 
${
\begin{array}{rrrrr}
x_1 & x_2 & x_3 & x_4\\ \hline
1 & 1 & 1 & 1\\
1 & 1 & 1 &-1\\
1 & 1 &-1 & 1\\
1 &-1 &-1 &-1\\
-1 & 1 &-1 &-1\\
-1 &-1 & 1 & 1\\
& & & \\
\end{array}
}$ & 
${
\begin{array}{rrrrr}
x_1 & x_2 & x_3 & x_4\\ \hline
 1 & 1 & 1 & 1\\
 1 & 1 & 1 &-1\\
 1 &-1 &-1 & 1\\
 1 &-1 &-1 &-1\\
-1 & 1 &-1 & 1\\
-1 & 1 &-1 &-1\\
-1 &-1 & 1 & 1
\end{array}
}$ \\ 
class  & affinely full-dim. & affinely full-dim. & subset FF\\
relation & none & none & $x_1x_2x_3 = 1$\\ 
$\max|\det M'M|$ & $2^8\cdot 3^2$ & $2^{10}\cdot 5$ & 
$2^{12}\cdot 3$ \\ \hline
$r$    & $8$ & $9$ & $10$\\
{\begin{tabular}{l}
$D$-optimal\\
design 
\end{tabular}} &
${
\begin{array}{rrrrr}
x_1 & x_2 & x_3 & x_4\\ \hline
 1 & 1 & 1 & 1\\
 1 & 1 & 1 &-1\\
 1 &-1 &-1 & 1\\
 1 &-1 &-1 &-1\\
-1 & 1 &-1 & 1\\
-1 & 1 &-1 &-1\\
-1 &-1 & 1 & 1\\
-1 &-1 & 1 &-1\\
& & & \\
& & & \\
\end{array}
}$ &
${
\begin{array}{rrrrr}
x_1 & x_2 & x_3 & x_4\\ \hline
1 & 1 & 1 & 1\\
1 & 1 & 1 &-1\\
 1 & 1 &-1 & 1\\
 1 &-1 & 1 & 1\\
 1 &-1 &-1 &-1\\
-1 & 1 & 1 & 1\\
-1 & 1 &-1 &-1\\
-1 &-1 & 1 &-1\\
-1 &-1 &-1 & 1\\
& & & \\
\end{array}
}$ &
${
\begin{array}{rrrrr}
x_1 & x_2 & x_3 & x_4\\ \hline
 1 & 1 & 1 & 1\\
 1 & 1 & 1 &-1\\
 1 & 1 &-1 & 1\\
 1 &-1 & 1 & 1\\
 1 &-1 &-1 &-1\\
-1 & 1 & 1 & 1\\
-1 & 1 & 1 &-1\\
-1 & 1 &-1 &-1\\
-1 &-1 & 1 &-1\\
-1 &-1 &-1 & 1
\end{array}
}$ \\ 
class  & regular FF & affinely full-dim. & affinely full-dim.\\
relation & $x_1x_2x_3 = 1$ & none & none\\ 
$\max|\det M'M|$ & 
 $2^{15}$ & $2^{12}\cdot 13$ & $2^{12}\cdot 3\cdot 7$ \\ \hline
\end{tabular}
\end{center}
\end{table}

\begin{table}[htbp]
\begin{center}
\caption{$D$-optimal designs for $5$ factors}
\label{tbl:non-saturated-s=5}
{\small
\begin{tabular}{lccc}\hline
$r$    & & $6$ & $7$\\
{\begin{tabular}{l}
$D$-optimal\\
design 
\end{tabular}}
& 
&
${
\begin{array}{rrrrr}
x_1 & x_2 & x_3 & x_4 & x_5\\ \hline
 1 & 1 & 1 & 1 & 1\\
 1 & 1 & 1 &-1 &-1\\
 1 & 1 &-1 & 1 &-1\\
 1 &-1 &-1 &-1 & 1\\
-1 & 1 &-1 &-1 & 1\\
-1 &-1 & 1 & 1 &-1\\
& & & & \\
\end{array}
}$ & 
${
\begin{array}{rrrrr}
x_1 & x_2 & x_3 & x_4 & x_5\\ \hline
 1 & 1 & 1 & 1 & 1\\
 1 & 1 & 1 &-1 &-1\\
 1 & 1 &-1 & 1 &-1\\
 1 &-1 & 1 &-1 & 1\\
 1 &-1 &-1 & 1 & 1\\
-1 & 1 &-1 &-1 & 1\\
-1 &-1 & 1 & 1 &-1\\
\end{array}
}$ \\
class & & affinely full-dim. & subset FF\\
relation & & none & $x_2x_3x_4x_5 = 1$\\ 
$\max|\det M'M|$ & & $2^{10}\cdot 5^2$ & $2^{16}$ \\ \hline
$r$    & $8$ & $9$ & $10$\\
{\begin{tabular}{l}
$D$-optimal\\
design 
\end{tabular}} & 
${
\begin{array}{rrrrr}
x_1 & x_2 & x_3 & x_4 & x_5\\ \hline
 1 & 1 & 1 & 1 & 1\\
 1 & 1 & 1 &-1 &-1\\
 1 &-1 &-1 & 1 & 1\\
 1 &-1 &-1 &-1 &-1\\
-1 & 1 &-1 & 1 &-1\\
-1 & 1 &-1 &-1 & 1\\
-1 &-1 & 1 & 1 &-1\\
-1 &-1 & 1 &-1 & 1\\
& & & & \\
& & & & \\
\end{array}
}$ & 
${
\begin{array}{rrrrr}
x_1 & x_2 & x_3 & x_4 & x_5\\ \hline
 1 & 1 & 1 & 1 & 1\\
 1 & 1 & 1 & 1 &-1\\
 1 & 1 & 1 &-1 &-1\\
 1 &-1 &-1 & 1 & 1\\
 1 &-1 &-1 &-1 &-1\\
-1 & 1 &-1 & 1 &-1\\
-1 & 1 &-1 &-1 & 1\\
-1 &-1 & 1 & 1 &-1\\
-1 &-1 & 1 &-1 & 1\\
& & & & \\
\end{array}
}$ & 
${
\begin{array}{rrrrr}
x_1 & x_2 & x_3 & x_4 & x_5\\ \hline
 1 & 1 & 1 & 1 & 1\\
 1 & 1 & 1 & 1 &-1\\
 1 & 1 & 1 &-1 & 1\\
 1 & 1 &-1 &-1 &-1\\
 1 &-1 &-1 & 1 & 1\\
 1 &-1 &-1 &-1 &-1\\
-1 & 1 &-1 & 1 & 1\\
-1 & 1 &-1 &-1 &-1\\
-1 &-1 & 1 & 1 &-1\\
-1 &-1 & 1 &-1 & 1
\end{array}
}$\\
class & regular FF & subset FF & affinely full-dim.\\
relation & $x_1x_2x_3 = x_1x_4x_5 = 1$ & $x_1x_2x_3 = 1$ & none\\ 
$\max|\det M'M|$ & $2^{18}$ & $2^{16}\cdot 7$ & $2^{14}\cdot 7^2$\\ \hline
\end{tabular}
}
\end{center}
\end{table}

We have not yet derived
any clear relation between the $D$-optimality and the affine full-dimensionality
like Conjecture \ref{conj:mod-8-rule} at present
from the results in Table \ref{tbl:non-saturated-s=4} and 
\ref{tbl:non-saturated-s=5}.

\section{Discussion}
We classify non-regular designs by
whether it is a proper subset of some regular fractional
factorial design (namely, subset fractional factorial design) 
or not (namely, affinely full-dimensional factorial design). 
We also give a geometrical interpretation of each class, i.e.,
the affinely full-dimensional factorial design is characterized 
as the design with the points which are not contained in any 
affine hyperplane. 
One justification of our definition 
is the simultaneous identifiability
of the parameters for the main effect model, which is one
of the most important concepts in the theory of designed 
experiments. Therefore, if the main purpose of the data analysis
is the estimation of the parameters for the main effect model, 
the strategy of considering the affinely full-dimensional factorial
designs is useful.

It is an interesting topic to investigate whether various ``good'' designs are
the affinely full-dimensional factorial designs or not. In this paper,
we consider this problem from the concept of $D$-optimality. In particular,
from the investigation of the $D$-optimal designs for the saturated cases,
interesting mod $8$ periodicity is suggested. Though the conjecture is not
proved at present, it holds for all the proved $D$-optimal designs 
under $100$ runs. We believe that the conjecture interests researchers
studying Hadamard maximum determinant problem.

It is also important to consider other optimal criteria such as $A$- 
or $E$-optimality. As an initial investigation, we have confirmed that
the $D$-optimal designs of saturated designs for $s = 4,5$ in Table 2 and 3
are also $A$-optimal and $E$-optimal designs.

\bibliographystyle{plain}

\end{document}